\documentclass[draft,jgrga]{agutex-1}
\newcommand{\car}{{$\mathrm{CO_2}$}}
 \newcommand{\dgr}{{$^\circ$}}
 \newcommand{\etal}{{et al.}}
\usepackage{graphicx}
 \setkeys{Gin}{draft=false}
\authorrunninghead{BHARDWAJ AND JAIN}

\titlerunninghead{ELECTRON ENERGY DEPOSITION IN CO$_2$}

\begin{document}

%%-------------------------------------------------------------------- %%
%
%  TITLE
%
%%-------------------------------------------------------------------- %%

\title{Monte Carlo model of electron energy degradation in a CO$_2$ 
atmosphere}
%
% e.g., \title{Terrestrial Ring Current:
% Origin, Formation and Decay $\alpha\beta\Gamma\Delta$}
% You may use \\ to break the title over several lines.

%% ------------------------------------------------------------------------ %%
%
%  AUTHORS AND AFFILIATIONS - 2 methods
%
%% ------------------------------------------------------------------------ %%

% Method 1 
% For three or fewer author/affiliation blocks, use \author{} and \affil{}

\author{Anil Bhardwaj and Sonal Kumar Jain}
\affil{Space Physics Laboratory, Vikram Sarabhai Space Centre, 
Trivandrum 695022, India \\ [30pt]
Published in Journal of Geophysical Research, Vol. \textit{114}, A11309 (2009), doi:10.1029/2009JA014298}

%\author{Sonal Kumar Jain}
%\affil{Space Physics Laboratory, Vikram Sarabhai Space Centre, 
%Trivandrum 695022, India}
%\affil{Department of Geography, Ohio State University,
%Columbus, Ohio, USA}

%\author{J. R. McConnell}
%\affil{Desert Research Institute, Division of Hydrologic Sciences,
%Reno, Nevada, USA}

% ---------------
% Method 2 
% For more than three author/affiliation blocks,
% use \author{\altaffilmark{}} and \altaffiltext{}
% \altaffilmark will produce footnote;
% matching altaffiltext will appear at bottom of page.
% May use \\ to start a new line.

% \authors{R. C. Bales, \altaffilmark{1}
% E. Mosley-Thompson, \altaffilmark{2}
% R. Williams, \altaffilmark{3}
% and J. R. McConnell\altaffilmark{4}}

% \altaffiltext{1}
% {Department of Hydrology and Water Resources, University of Arizona,
% Tucson, Arizona, USA.}
%
% \altaffiltext{2}{Department of Geography, Ohio State University,
% Columbus, Ohio, USA.}
%
% \altaffiltext{3}{Department of Space Sciences, University of Michigan,
% Ann Arbor, Michigan, USA.}
%
% \altaffiltext{4}{Desert Research Institute, Division of Hydrologic Sciences,
% Reno, Nevada, USA.}

%%--------------------------------------------------------------------- %%
%
%  ABSTRACT
%
%%-------------------------------------------------------------------- %%

% >> Do NOT include any \begin...\end commands within
% >> the body of the abstract.

\begin{abstract}
A Monte Carlo model has been developed to study the degradation of 
$\le$1000 eV  electrons in an atmosphere of \car, which is one of the most 
abundant species in Mars' and Venus' atmospheres. The 
e-\car\ cross sections are presented in an assembled 
set along with their analytical representations. 
Monte Carlo simulations are carried out at several energies to 
calculate the ``yield spectra'', which embodied all the information 
related to electron degradation process and can be used to calculate 
``yield'' (or population) for any inelastic process. The numerical yield 
spectra have been fitted analytically resulting in an analytical yield 
spectra (AYS). We have calculated the mean energy per ion pair and 
efficiencies for various inelastic processes, including the double 
and dissociative double ionization of \car\ and  negative ion 
formation. The energy distribution of the secondary electrons produced per 
incident electron is also presented at few incident energies. 
The mean energy per ion pair for \car\ is 37.5 (35.8) eV at 200 (1000) eV, 
compared to experimental value 32.7 eV at high energies. 
Ionization is the dominant loss process at energies above 50 eV with 
contribution of $\sim$50\%. Among the excitation processes, 13.6 eV and 
12.4 eV states are the dominant loss processes consuming $\sim$28\% 
energy above 200 eV. Around and below ionization threshold, 13.6 eV, 12.4 
eV, and 11.1 eV, followed by 8.6 eV and 9.3 eV
excitation states are important loss processes, while below 10 eV 
vibrational excitation dominates.
\end{abstract}

%%--------------------------------------------------------------------- %%
%
%  BEGIN ARTICLE
%
%%------------------------------------------------------------------- %%

% The body of the article must start with a \begin{article} command
%
% \end{article} must follow the references section, before the figures
%  and tables.

\begin{article}

%% ------------------------------------------------------------------------ %%
%
%  TEXT
%
%% ------------------------------------------------------------------------ %%
%%modifide on 22-09-08 introduction and Monte Carlo section..
\section{Introduction}
Photoionization is the main source of electrons and ions in the dayside 
upper atmosphere of planets. Photoelectrons, generated due to 
photoionization process, can  have enough  kinetic energy to ionize the 
atmospheric constituents and produce secondary electrons. Similarly, 
energetic electrons precipitating along the magnetic field lines into the 
auroral atmosphere of planets can ionize the medium producing secondary
electrons.
Besides ionization, the electron energy is lost in excitation, 
attachment, and dissociation. Hence, the study of  
electron energy deposition in atmosphere is an important 
aspect in understanding processes like aurora, dayglow, nightglow
[\textit{e.g., Bhardwaj and Gladstone}, 2000; \textit{ Fox et al.}, 2008].
To model the electron energy degradation in an atmosphere one has 
to first compile cross sections for various loss processes, and 
then develop an electron energy apportionment method, 
which will distribute the electron energy among different loss 
channels. 

The study of the electron energy degradation in  \car\ is of fundamental 
interest in various fields of science.  
\car\ is one of the most important molecules in our solar system. 
It comprise more than 90\% of the atmospheres of Venus and Mars. 
It is also used in lasers, gaseous discharge or low power 
plasma device. Electron energy degradation in \car\ gas has 
important applications to Mars and Venus. Earlier results from  Mariner 
satellites and Mars 3 and Mars 4 spacecrafts have confirmed the presence 
of an ionosphere on Mars, and also detected various emission features on 
Mars [\textit{e.g., Barth et al.}, 1971; \textit{Dementyeva et al.}, 1972],
which have been studied in detail by recent SPICAM ultraviolet 
spectrometer observations aboard  Mars Express [\emph{e.g., Bertaux 
et al.}, 2006; \emph{Leblanc et al.}, 2006]. Emissions 
from Venus have been studied quite extensively by Pioneer Venus 
[\textit{e.g., Fox and Bougher}, 1991] and by the ongoing Venus Express
[\textit{e.g., Bertaux et al}, 2007]. Electron impact excitation and 
dissociative excitation of \car\ are the key processes in the production 
of several emissions on Mars and Venus.

In this paper we present a Monte Carlo model which describes the energy 
degradation of $\leq$1000 eV electrons in an atmosphere of \car. 
Earlier studies of electron degradation in \car\ have been carried out
by \textit{Sawada et al.} [1972], \textit{Green et al.} [1977], and 
\textit{Fox and Dalgarno} [1979]. Monte Carlo methods are  class of 
numerical methods based on stochastic technique. Though it is 
time consuming, but due to its probabilistic nature, it is an excellent 
technique for studying the energy degradation of particles, provided 
sufficient sample size is taken. Hence, Monte Carlo methods have been 
widely used in problems dealing with energetic particle degradation in 
gases and in applications to the planetary atmospheres [\textit{e.g.}, 
\emph{Cicerone and Bowhill}, 1971; \emph{Ashihara}, 1978; 
\emph{Green et al.}, 1977, 1985; \emph{Singhal et al.}, 1980; 
\emph{Singhal and Green}, 1981; \emph{Singhal and Bhardwaj}, 1991; 
\emph{Bhardwaj and Singhal}, 1993; \emph{Michael and Bhardwaj}, 
2000; \emph{Bhardwaj and Michael}, 1999a, b; \textit{Shematovich et 
al.}, 2008].  

In section 2, we present a compilation of all the e-\car\ loss processes 
cross sections available up to the present date and fitted them with a 
simple analytical form. These analytically fitted cross sections can be 
easily used in the Monte Carlo model, which is presented in section 3. 
The output of the Monte Carlo simulation is employed to  generate a 
``yield spectrum,'' which is presented in section 4.
The concept of the yield spectrum was first introduced by 
\emph{Green et al.} [1977] and further developed by many workers 
[\textit{e.g.}, \emph{Green and Singhal}, 1979;  
\emph{Singhal and Green}, 1981; \textit{Singhal and Haider}, 1984; 
\emph{Green et al.}, 1985; \emph{Singhal and Bhardwaj}, 1991; 
\emph{Bhardwaj and Singhal}, 1993; \emph{Bhardwaj and Michael}, 1999a].
The yield spectra embodied the information about the electron
degradation processes and can be used to calculate ``yield'' for
any inelastic event. The numerical yield spectrum is represented in
an analytical form resulting in an analytical yield spectrum (AYS). 
The AYS and its comparison with the numerical yield spectrum  
is also presented in the section 4.  
In sections 5 and 7, we present the calculated mean energy per ion pair 
and efficiencies for inelastic processes, respectively, using AYS and 
compare them with that obtained by using numerical yield spectra. The 
energy distribution of secondary and tertiary electrons produced during 
ionization events is presented in section 6. Summary of the paper is 
presented in section 8.

\section{Cross sections}
\subsection{Total}
The laboratory measured total scattering cross section (TCS) is  
available between 0.1 eV and 5000 eV. The TCS for e-CO$_2$ collision 
has been measured by several authors in different energy ranges --
\emph{Ferch et al.} [1981] in the energy range 0.007-4.5 eV, 
\emph{Buckman et al.} [1987] 0.1-5 eV, \emph{Szmytkowski et al.} [1987] 
0.5-3000 eV , \emph{Kimura et al.} [1997] 0.8-500 eV, 
\emph{Kwan et al.} [1983] 1-500 eV, and
\emph{Garcia and Manero} [1996] 400-5000 eV.
At low energies, the TCS of \emph{Szmytkowski et al.} [1987],
\emph{Buckman et al.} [1987], and \emph{Ferch et al.} [1981]
are in agreement to  within 10\%. Recently, \emph{Zecca et al.} [2002]
have determined the best value of TCS. In the lowest energy range 
($<$1 eV) \emph{Zecca et al.} [2002] 
adopted the experimental data of \emph{Ferch et al.} [1981] and 
\emph{Buckman et al.} [1987], which are in good  agreement
with each other. In the 1-1000 eV energy range, \emph{Zecca et al.}
[2002] averaged the cross sections obtained
by \emph{Szmytkowski et al.} [1987], \emph{Kimura et al.}, [1997]
and \emph{Kwan et al.} [1983], with equal weight, to obtain the
recommended values, which are in good agreement with
\emph{Garcia and Manero} [1996] at higher ($>$400 eV) energies. 
In his review, \emph{Itikawa} [2002] has recommended the TCS of  
\emph{Zecca et al.} [2002].
The TCS reaches a maximum value of $60\times10^{-16}$ cm$^2$
 at 0.1 eV [\emph{Ferch \etal}, 1981; \emph{Buckman \etal}, 1987], it then 
goes through a minimum of $5.5\times10^{-16}$ cm$^2$ at 1.9 eV
[\emph{Szmytkowski et al.} 1987]. At lower energies a resonant structure is
present $\sim$3.8 eV.

\subsection{Elastic}
\subsubsection{Differential elastic}
The differential elastic scattering cross section (DCS) for e-CO$_2$  
collision has been measured by many authors [cf. review by 
\textit{Itikawa}, 2002; \textit{Karwasz et al.}, 2001].
 
In the 1-4 eV energy, the DCS values of \emph{Gibson et al.} [1999] 
and \emph{Tanaka et al.} [1998] are in good agreement at forward 
angles ($\leq$50\dgr), however at larger angles they differ by 
20-30\%. Overall, at most of the energies there are good agreement 
in shape between these two DCS.
At 30, 40, and 50 eV, the DCS measurements of 
\emph{Gibson et al.} [1999], {Kanik et al.} [1989], and 
\emph{Tanaka et al.} [1998] are in reasonable accord, within the 
uncertainties of each measurement, and at 50 eV the DCS of 
\emph{Gibson et al.} [1999] and {Register et al.} [1980] are 
consistent. At 100 eV, the measured DCS values of \emph{Iga et al.} [1999] 
are in good agreement with \emph{Kanik et al.} [1989] and 
\emph{Tanaka et al.} [1998]. 

We have taken the DCS values from \emph{Tanaka et al.} [1998] in the 1-100
eV range, however values at 40, 50, 70, 80, and 90 eV are taken from 
\emph{Kanik et al.} [1989] which agree well with the cross section of 
\emph{Tanaka et al.} [1998] in the entire energy range. The DCS values in 
200-400 eV range are taken from \emph{Iga et al.} [1999], and those in 
500-1000 eV are taken from \emph{Iga et al.} [1984]. In Table 1, we 
present the DCS values used in this work.

\subsubsection{Total elastic}
Based on the DCS measured by \emph{Register et al.} [1980], 
\emph{Tanaka et al.} [1998]  and {Gibson et al.} [1999],
\emph{Buckman et al.} [2002] have determined 
the total elastic cross section in 1-100 eV range with an 
estimated uncertainty of $\pm30\%$. \emph{Shirai et al.} [2001] have 
reported the recommended  elastic cross section up to 1000 eV by 
considering the  beam data of \emph{Iga et al.} [1999].
\emph{Itikawa} [2002] has recommended the elastic cross section of 
\emph{Buckman et al.} [2002] in the energy range 1-60 eV, and 
\emph{Shirai et al.} [2001] in the energy range 100-1000 eV. The two 
data sets merge smoothly .

We have taken total elastic cross section as recommended by 
\emph{Itikawa} [2002]. The total elastic cross section is fitted using
the semi-empirical formula [\emph{Bhardwaj and Michael}, 1999a]:

\begin{eqnarray}
\sigma(E)  =  \frac{1}{A_1+B_1E}+\frac{1}{A_2+B_2E}+  
               \frac{2}{E}\frac{\sqrt{A_1A_2}}{A_2B_1-A_1B_2} %\\ \nonumber
               \ln\frac{(1+B_1E/A_1)}{(1+B_2E/A_2)},     %eq 1
\end{eqnarray}
where $A_1, B_1, A_2,$ and $B_2$ are the fitting parameters,
whose values are $8.090\times10^{-16}$ $\mathrm{\AA}^{-2}$,
$2.184\times10^{-2}$ $\mathrm{\AA}^{-2}$ keV, $0.92$
$\mathrm{\AA}^{-2}$ and $5.0\times10^{-4}$
$\mathrm{\AA}^{-2}$ keV, respectively, and $E$ is the energy of
the electron in eV. Lower limit of fit is 30 eV, and fitted cross section 
is shown in Figure 1. At energies below 30 eV it is difficult to fit the 
cross section using above equation due to resonance structure present at 
low energies ($\sim$4 eV), and hence these values are fed numerically in 
the Monte Carlo model.

\subsection{Dissociative electron attachment}
The dissociative attachment process in e-CO$_2$ collisions, which
mainly occurs at energies $<$12 eV, leads to the formation of negative
ions O$^{-}$, O$_2^-$, and  C$^{-}$. \emph{Rapp and Briglia} [1965]
measured absolute values of the total cross section for the production of 
negative ions from CO$_2$. \emph{Orient and Srivastava} [1983] 
obtained the cross section for the production of O$^{-}$ ions and 
showed that it is the dominant anion. 
Their values are in agreement with those of
\emph{Rapp and Briglia} [1965] within the uncertainty of the
cross sections ($\pm20\%$) and the energy scale ($\pm0.1$ eV).
\emph{Spence and Schulz} [1974] measured the cross sections for the
production of C$^{-}$ and O$_2^-$ ions. The cross section for O$_2^-$ 
production has two peaks of the order of $10^{-24}$ cm$^2$ at 11.3 and 
12.9 eV, while cross section for C$^{-}$ production has three peaks 
with the largest value of  $\sim$$2\times 10^{-21}$ cm$^2$.
The cross sections for O$_2^-$ and C$^{-}$ are small compared to 
that of O$^{-}$, and hence are not considered in our study.

We have adopted the cross section values of \emph{Rapp and Briglia}
[1965] for the production of O$^{-}$ ions from CO$_2$. The cross
section shows a double-peak structure -- peaks at 4.1 and 8.3 eV, with the 
later peak value ($4.28\times 10^{-19}$ cm$^2$) about 2.5 times the value 
of the former peak. The cross section for each peak has been fitted 
with the following analytical form [\textit{Bhardwaj and Michael}, 1999a]:
 
\begin{equation}
        \sigma(E)=\frac{Ae^{t}/U}{(1+e^{t})^2},      %eq 2
\end{equation}
Here $t=(E-W_p)/U$, where $W_p$ is the energy at the peak. The values
of the overall normalization parameter $A$ and the effective width
parameter $U$ for each of the peaks along with the
parameter $W_p$ and threshold energy $W_{th}$ are presented in Table 2. 
The fitted cross sections along with laboratory measurements are given in 
Figure 2.

\subsection{Ionization}
The ionization and dissociative ionization of CO$_2$ by electron
impact produce singly and doubly ionized ions (CO$_2^+$, CO$^+$, 
C$^+$, O$^+$, C$^{++}$, O$^{++}$, and CO$_2^{++}$). The cross 
sections for these processes have been reported by \emph{Rapp and 
Englander-Golden} [1965], \emph{Shyn and Sharp} [1979], \emph{Orient 
and Srivastava} [1987], \emph{Tian and Vidal} [1998], and \textit{Straub et
al.} [1996]. Recently, \emph{McConkey et al.} [2008] have reviewed the 
electron impact dissociation cross sections for \car.
For the total ionization cross section, measurements of 
\emph{Orient and Srivastava} [1987], \emph{Tian and Vidal} [1998],
and \emph{Straub et al.} [1996] are within the error limits with
values of \emph{Rapp and Englander-Golden} [1965] upto 1000 eV, and 
with the data of \emph{Shyn and Sharp} [1979] in the energy
range 50-400 eV. 
\emph{Tian and Vidal} [1998] have also measured the cross sections for 
double and triple ionization of \car\ due to electron impact. 
After a survey of the available experimental data,
\emph{Lindsay and Mangan} [2002] suggested recommended values 
of ionization cross section. Their partial cross sections are based on 
measurement of \emph{Straub et al.} [1996]. For total ionization cross 
section below 25 eV, \emph{Lindsay and Mangan} [2002] adopted the values of
\emph{Rapp and Englander-Golden} [1965]. At energies above 25 eV, they
reported uncertainties of 5\%  for the partial cross sections 
for the production of CO$_2^+$, CO$^+$, C$^+$, O$^+$, and the total 
ionization cross section. The cross sections at energies below 25 eV
have uncertainties of 7\%. 
There are also uncertainties in appearance energies of fragmented ions 
CO$^+$, C$^+$, O$^+$, C$^{++}$, and O$^{++}$. We have taken the appearance
energies for the fragmented ions from \emph{Itikawa} [2002].
   
We have used the dissociative and direct ionization cross sections 
recommended by \emph{Lindsay and Mangan} [2002] [cf. \emph{Itikawa},
2002; \textit{McConkey et al.}, 2008]. The CO$_2^+$ ion can be produced 
in four excited states, viz., X$^2\Pi_g$,  A$^2\Pi_u$, 
B$^2\Sigma_u^+$, and C$^2\Sigma_g^+$. Cross sections for A$^2\Pi_u$ and 
B$^2\Sigma_u^+$ states have been taken from \textit{Itikawa} [2002], 
while the cross sections for X$^2\Pi_g$ and C$^2\Sigma_g^+$ states have 
been taken from \textit{Jackman et al.} [1977]. For double ionization, 
cross sections of (CO$^+$,O$^+$), (C$^+$,O$^+$), and (O$^+$,O$^+$) 
production have been taken from \emph{Tian and Vidal} [1998] up to 600 eV;
these cross sections have not been added in the total ionization cross 
section because they are already accounted in the cross sections for the 
formation of CO$^+$, C$^+$, and O$^+$ ions. All these cross 
sections have been fitted using the analytical expression  
[\emph{Jackman et al.}, 1977; \textit{Bhardwaj and Michael}, 1999a]. 

\begin{equation}
\sigma(E)=A\Gamma \left[arctan\frac{(T_M-T_0)}{\Gamma}
        + arctan\left(\frac{T_0}{\Gamma} \right)\right],\nonumber%}
\end{equation}                  %eq 3
where
$$
A(E)=\left[\frac{K}{E+K_B}\right]\ln\left[\frac{E}{J}+J_B+\frac{J_C}{E}
\right];
$$

$$ \Gamma(E)=\Gamma_S\left[\frac{E}{E+\Gamma_B}\right]; $$

$$
T_0(E)=T_S-\left[\frac{T_A}{E+T_B}\right]; \\\ T_M=\frac{E-I}{2}.
$$
Here $E$ is the incident energy in eV, $I$ is the fitting ionization
potential in the eV, which is generally close to the threshold
potential (W$_{th}$), and $\sigma$ is in units of $10^{-16} $ cm$^2$.
This form gives the asymptotic behavior $\sigma(E)\propto E^{-1}\ln E$ at
high energies, which is expected from the Born approximation. The
fitting parameters are presented in the Table 3. The fitted cross sections 
for single and double ionization are shown in Figures 3 and 4, 
respectively.

\subsection{Excitation cross sections}
\subsubsection{Vibrational excitation}
CO$_2$ is a linear triatomic molecule, which has  three normal modes of
vibration, \textit{i.e.}, a bending mode (0 n 0), a symmetric stretching 
mode (n 0 0), and an asymmetric stretching mode (0 0 n), with excitation 
energy 83 meV, 172 meV, and 291 meV, respectively 
[\textit{Kochem et al.}, 1985]. Infrared
active (010) bending and (001) asymmetric stretching modes in the 
near-to-threshold region follow the Born approximation. Moreover, the 
structure near the threshold of vibration excitation in CO$_2$ has been
investigated by \textit{Kochem et al.} [1985], 
vibrationally inelastic DCS above 4 eV impact energies
have been measured by \textit{Register et al.} [1980] for scattering
angles $10^\circ-140^\circ$ and impact energies of 4, 10, 20,
and 50 eV, and by \textit{Johnstone et al.} [1995] for only one scattering 
angle ($20^\circ$) in the energy region 1 to 7.5 eV. 
\textit{Nakamura} [1995] determined the vibrational cross section 
using swarm experiment. 
\textit{Kitajima et al.} [2001] made  measurements of DCS for the 
electron impact excitation of CO$_2$ for (010), (100), (001), and (020)
vibrational modes over the scattering  angles $20^\circ-130^\circ$ and 
energy range 1.5-30 eV (except at 4 eV where the smallest angle was 
extended upto $10^\circ$), and assigned an uncertainty of 30\% to 
their measurements.
Their DCS is consistent with the results of previous beam-type 
measurements. 
\textit{Itikawa} [2002] has extrapolated the DCS of 
\textit{Kitajima et al.} [2001] to obtain the total vibration cross 
sections for three modes, which are presented in Figure 1.

In our studies we have taken cross sections for the three fundamental 
vibrational modes (010), (100), and (001) from \textit{Itikawa} 
[2002]. There are other modes also but their cross sections 
are small compare to these three fundamental modes. 
                
\subsubsection{Electronic excitation}
There are several features in the optical and electron
scattering spectrum of \car\ in the energy loss range between 7 and
11 eV (\emph{Herzberg}, 1966; \emph{Rabalais et al.}, 1971;
\emph{Hall et al.}, 1973). Except for Rydberg
states, there is still no definite consensus about structure and
assignment of the excited electronic states of \car. 
In the energy loss spectra of \car, \emph{Green et al.} [2002]
have found four clearly distinct peaks at 10.98, 11.05, 11.16, and 
11.40 eV, with an uncertainty of 30\% in their results. 
\emph{Itikawa} [2002] in his review paper has recommended the DCS of 
\emph{Green et al.} [2002], for the excitation of the 10.8-11.5 eV 
energy loss states. Recently, \emph{Kawahara et al.} [2008] have
given the integral cross section for electronic states $^1\Sigma_u^+$
and $^1\Pi_u$ of \car, based on the DCS measurement of 
\emph{Green et al.} [2002] in the energy range 20-200 eV.
                      
Theoretical calculations of electronic structure have also been made
by several authors [\emph{Nakatsuji}, 1983; \emph{Spielfiedel et al.},
1992; \emph{Buenker et al.}, 2000; \emph{Lee et al.}, 1999].
Using distorted-wave method, \emph{Lee and McKoy} [1983] calculated
the cross section for the excitation of eight low lying-states.
But there is not much agreement among these calculations.
In summary, there is still a need for a detailed study of excitation of 
electronic states of \car\ by electron impact.
   
We have taken the empirical cross sections of \emph{Jackman et al.}
[1977] for the electronic states of \car. These cross sections have been
obtained using equation: 

\begin{equation}
 \sigma(E)=\frac{(q_0F)}{W^2}\left[1-\left(\frac{W}{E}\right)^\alpha
 \right]^\beta \left[\frac{W}{E}\right]^\Omega      %eq 4
\end{equation}
where $q_0=4\pi a_0R^2$ and has the value $6.512\times10^{-14}$ eV$^2$
cm$^2$. The fitting parameter are given in Table 4.
The parameters for the two states 12.4 and 13.6 eV, that corresponds to 
Cameron band of CO [cf. \emph{Sawada et al.}, 1972] have been modified. 
The peak cross section of their sum is $2.40\times 10^{-16}$ cm$^{2}$ 
at 80 eV [\emph{Erdman and Zipf}, 1983].

\subsection{Emission}
Electron impact dissociation and ionization of CO$_2$ can result in 
the production of excited fragments of CO, O, and
CO$_2$ in the neutral and ionized states, resulting in the emissions in
the ultraviolet region. These emissions are important for understanding 
phenomena like aurora, dayglow that occur in the atmospheres of Mars, 
Venus, and \car-containing  atmospheres.  
The strong band systems observed on Mars are Fox-Duffendack-Barker bands 
($A^2\Pi_u \rightarrow X^2\Pi_g$) and ultraviolet doublet 
($B^2\Sigma_u^+\rightarrow X^2\Pi_g$) of CO$_2^+$, and Cameron bands 
($a^3\Pi \rightarrow X^1\Sigma^{+}$) of CO [\emph{Ajello}, 1971; 
\emph{Barth et al.}, 1971; \emph{Bertaux et al.}, 2006; \emph{Leblanc et 
al.}, 2006]. \emph{Ajello} [1971] measured  the emission cross sections 
for the $A^2\Pi_u \rightarrow X^2\Pi_g$ and 
$B^2\Sigma_u^+\rightarrow X^2\Pi_g$ bands of CO$_2^+$ from threshold to 
300 eV. He also measured cross sections for the excitation of the fourth 
positive system of CO ($A^1\Pi\rightarrow X^1\Sigma^{+}$), the first 
negative system of CO$^+$ ($B^2\Sigma^{+}\rightarrow X^2\Sigma^{+}$) and 
several atomic multipletes of carbon and oxygen produced from dissociative
excitation of \car.

\subsubsection{Emission from CO$_2^+$}
\emph{McConkey et al.} [1968], \emph{Ajello} [1971], and \emph{Tsurubuchi
and Iwai} [1974] have detected emissions corresponding to the
following transitions: 

$$
        A^2\Pi_u \rightarrow X^2\Pi_g
     \quad at\ 293.6 - 438.4\ nm
$$
and
$$
     B^2\Sigma_u^+\rightarrow X^2\Pi_g
    \quad at\ 218.9 - 226.8\ nm
$$
The peak value of cross sections measured by the
three groups for the above transitions are in good agreement with
each other. These emissions are well known in the Mars upper atmosphere. 
Both the ground and excited states of CO$_2^+$ are known
to be linear [\emph{Herzberg}, 1966]. The cross section of \emph{Ajello}
[1971] has too steep an energy dependence near threshold  compared
to \emph{McConkey et al.} [1968] and \emph{Tsurubuchi and Iwai}
[1974]. In his review, \emph{Itikawa} [2002] recommended the 
cross sections of \emph{Tsurubuchi and Iwai} [1974], 
for which the peak values are $(8.0\pm2.0)\times10^{-17}$ cm$^2$ at
160 eV for the $ A - X $ transition, and $(4.7\pm1.2)\times10^{-17}$
cm$^2$ for the $B - X$ transition. We have taken the cross 
sections for $ A - X $ and $B - X$ emissions of CO$_2^+$ from 
\emph{Itikawa} [2002]. These cross sections have been fitted using 
equation (3). The fitting parameters are given in Table 3, and fitted 
cross sections in Figure 3.

\subsubsection{Emission from CO$^+$}
Only \emph{Ajello} [1971] has measured the cross section for the emission 
of first negative system ($B^2\Sigma^+ \rightarrow X^2\Sigma^+$) of CO$^+$.
The cross section exhibits an appearance potential of 25.11 eV, and the 
peak value of cross section is $1.9\times 10^{-18}$  cm$^2$ around 100 eV.
The cross section for the excitation of the first negative system
of CO$^+$ from electron impact on \car\ is about a factor of 25 less than 
for excitation of the
same system from CO [\emph{Ajello}, 1971]. We have adopted the cross 
section of \emph{Ajello} [1971], which has been 
fitted analytically using equation (4); the fitting parameters are given 
in Table 4. Figure 3 shows the fitted cross section along with
experimental cross section.

\subsubsection{Emission from CO}
Cross sections for the production of Cameron band system 
($a^3\Pi \rightarrow X^1\Sigma^+$) and fourth 
positive system ($A^1\Pi \rightarrow X^1\Sigma^+$) of CO have been 
measured by \emph{Ajello} [1971]. The emission cross section for the 
fourth positive system is very weak and Ajello 
could not measure the cross section near threshold (13.48 eV). For the
Cameron band system, \emph{Ajello} [1971] reported relative magnitudes of
the cross section for the (0, 1) band at 215.8 nm. The upper state 
($a^3\Pi$) of Cameron emission is metastable and has a long radiative 
lifetime ($\sim$3 ms) [\emph{Gilijamse et al.}, 2007], and  
kinetic energies of the CO($a^3\Pi$) fragments are in the range of 
0--1.2 eV [\emph{Freund}, 1971]. 
\emph{Erdman and Zipf} [1983] measured the total cross
section for CO ($a^3\Pi \rightarrow X^1\Sigma^+$) electronic transition.
They estimated the absolute magnitude of total Cameron band emission 
cross section of $2.4\times10^{-16}$ cm$^2$ at 80 eV. The Cameron band is
the brightest emission feature in the UV dayglows of both Mars and Venus 
as well as an important emission in \car-containing atmospheres, 
\textit{e.g.} comets.

\subsubsection{Emission from O and C}
Both \textit{Ajello} [1971] and \textit{Mumma et al.} [1972] have reported 
cross section for the emission of the  O 130.4 nm triplet from  electron 
impact on \car, but the measurements are not consistent with each other. 
There are many other atomic emissions produced in e-\car\ collisions, but 
they have very small cross sections [cf. \textit{van der Burgt et al.}, 
1989]. \emph{Kanik et al.} [1993] have reported the emission cross 
sections for O, O$^+$, C, C$^+$, CO, and CO$^+$ in the
wavelength region 40 - 125 nm. All the cross sections of \emph
{Kanik et al.} [1993] are less than $10^{-18}$ cm$^2$. 
We have adopted the O I and C I production cross sections of 
\textit{Jackman et al.} [1977]. 

\section{Monte Carlo Model}
The transport of radiation is a natural stochastic process that is 
amenable to the Monte Carlo method due to its probabilistic nature. 
In the Monte Carlo simulation, modeling of an inherently stochastic 
system is carried out by artificial random sampling.
In the present work we have developed a Monte Carlo model to  simulate 
the local degradation of 1-1000 eV electrons in an atmosphere of CO$_2$ 
gas. The energy bin size is taken as 1 eV throughout the energy range.
In the simulation we have considered elastic 
scattering between electrons and neutral \car\ molecules, and various 
inelastic processes like ionization, excitation, attachment,  
dissociation, etc; the cross sections for these processes are described 
in section 2. Figure 5 illustrates how an individual electron is treated 
in the Monte Carlo simulation.

The initial energy $E_0$ of the electron is fixed at the beginning of
the simulation and the direction of movement of the electron ($\theta,\
\phi$) is decided with the help of two random numbers $R_1$
and $R_2$ [random numbers are uniformly distributed in the range (0, 1)] as

\begin{equation}                             
\theta=\cos^{-1}(1-2R_1),           %eq 5
\end{equation}
\begin{equation}                       
\phi=2\pi R_2.                      %eq 6
\end{equation}
The distance to next collision is calculated from
\begin{equation}
S = -\log(1-R_3)/n\sigma_T,         %eq 7
\end{equation}
where $R_3$ is a random number, $n$ is the number density of the
neutral target species (taken as $1\times10^{10}$ cm$^{-3}$), and 
$\sigma_T$ is the total (elastic + inelastic) electron impact collision 
cross section. After generating a new random number $R_4$, the probability
of elastic collision $P_{el}=\sigma_{el}/\sigma_T$ is calculated. 
if $P_{el} > R_4$, elastic collision takes place. if $P_{el}\leq R_4$, 
the inelastic event takes place, and in this case we further test for 
the type of inelastic event that has taken place with the help of 
another random number.

For elastic scattering the energy loss is calculated as 

\begin{equation}
\bigtriangleup E=\frac{m^2v^2}{m+M}-\frac{m^2vV_1\cos\delta}{m+M},
\end{equation}                      %eq 8
$$
V_1=v\left[\frac{m\cos\delta}{m+M}+\frac{[M^2+m^2(\cos\delta-1)]^{1/2}}
{m+M}\right].
$$
Here $\delta$ is the scattering angle in the laboratory frame, $v$ and
$m$ are the velocity and mass, respectively, of the electron, and $M$
is the mass of the target particle. Differential elastic cross sections
(discussed in section 2.2.1) are used to obtain the scattering
angle $\delta$. Differential cross sections are fed numerically in
the Monte Carlo model at 28 unequally spaced energy points (1.5, 2, 3,
3.8, 4, 5, 6, 6.5, 7, 8, 9, 10, 15, 20, 30, 40, 50, 60, 70, 80, 90,
100, 200, 300, 400, 500, 800, and 1000 eV) and at 20 scattering angles
($0^\circ$, $5^\circ$, $10^\circ$, $15^\circ$, $20^\circ$, $30^\circ$,
$40^\circ$, $50^\circ$, $60^\circ$, $70^\circ$, $80^\circ$, $90^\circ$,
$100^\circ$, $110^\circ$, $120^\circ$, $130^\circ$, $135^\circ$,
$150^\circ$, $165^\circ$, and $180^\circ$). At intermediate energies
and angular points the values are obtained through linear
interpolation. The energy $\bigtriangleup E$ is subtracted from the
energy of the test particle. After the collision, the deflection 
angle relative to the direction ($\theta,\phi$) is obtained as 

$$
\cos\theta^{''}=\cos\theta\cos\theta^{'}-\sin\theta\sin\theta^{'}
\cos\phi^{'} ,                    %eq 9
$$
\begin{eqnarray}
\cos\phi^{''}=(\cos\theta\cos\phi\sin\theta^{'}
            \sin\phi^{'}-\sin\phi\sin\theta^{'}\sin\phi^{'} % \\ \nonumber
           +\sin\theta\cos\phi\cos\theta^{'})/\sin\theta^{''},
\end{eqnarray}
\begin{eqnarray*}
\sin\phi^{''}=(\cos\theta\cos\phi\sin\theta^{'}\cos\phi^{'}
            -\cos\phi\sin\theta^{'}\sin\phi^{'} % \\ \nonumber
            +\sin\theta\sin\phi\cos\theta^{'})/\sin\theta^{''}.
\end{eqnarray*}
Here $\theta^{'}$, $\phi^{'}$ are the scattering angles.

In the case of an inelastic collision, the next step is to find whether 
the event is ionization or any of the other type of inelastic 
collision. If the collision is an ionization event, a secondary 
electron is produced. The energy of the secondary electron $T$ is 
calculated with the help of a random number $R$ as [\textit{Bhardwaj and 
Michael}, 1999a]

\begin{equation}
  T=\frac{\Gamma_S\ E_v}{E_v+\Gamma_B}[\tan(RK_1+(R-1)K_2)]+T_S
     -\left[\frac{T_A}{E_v+T_B}\right],          %eq 10
\end{equation}
where
$$
K_1 = \tan^{-1}\left\{\left[\frac{(E_v-I)}{2}-T_S
    +\frac{T_A}{(E_v+T_B)}\right]
    /\frac{\Gamma_S\ E_v}{(E_v+\Gamma_B)}\right\},
$$
$$
K_2 = \tan^{-1}\left\{\left[T_S
     -\frac{T_A}{(E_v+T_B)}\right]
     /\frac{\Gamma_S\ E_v}{(E_v+\Gamma_B)}\right\}.
$$
Here $E_v$ is the energy of the incident primary electron before the
ionization event. $\Gamma_S$, $\Gamma_A$, $T_A$, $T_B$, and $T_S$ are
the fitting parameters, and $I$ is the ionization threshold. The
values of these parameters are given in Table 3.
If the energy of secondary electron, produced in the ionization event,
is more than the lowest cutoff energy (which is 1 eV in our 
simulation) then it is also tracked in a same manner as the primary 
electron (cf. Figure 5). The secondary electrons can also cause 
ionization, producing tertiary electrons, which are treated in 
a similar way as secondary electrons. In the Monte Carlo simulation we 
also follow tertiary and subsequent electrons.  
The number of secondary, tertiary, and subsequent electrons produced 
during the ionization events are stored in the appropriate energy bins. 
After the type of collision event has been decided, the appropriate 
energy is subtracted from the energy of the particle. All the 
collision events are recorded in the appropriate energy bins 
corresponding to the energy of the electron at the time of collision. 
The history (track view) of a particle with each interaction event is 
traced until  the electron energy falls below an assigned cutoff value, 
which is 1 eV. The sample size in the present study 
is $10^6$ particles for each simulation.

\section{Yield Spectra}
When all the sampled electrons have been degraded, we get a two
dimensional yield spectrum, which is a function of the spectral
energy $E$ and incident primary electron energy $E_0$, defined as
[\emph{Green et al.}, 1977]:

\begin{equation}
      U(E,E_0)=\frac{N(E)}{\bigtriangleup E}, %eq 11
\end{equation}
where $N(E)$ is the number of inelastic collision events for which the
spectral energy of the electron is between $E$ and $E+\bigtriangleup E$,
where $\bigtriangleup E$ is the energy bin width, which is 1 eV in our
model. This yield spectrum is related to the degradation spectrum or
equilibrium flux $f(E,E_0)$ of \emph{Spencer and Fano} [1954] by the
equation

\begin{equation}
      U(E,E_0)=\sigma_T(E)f(E,E_0),            %eq 12
\end{equation}
where $\sigma_T$ is the total inelastic collision cross section.

The analytical yield spectrum $U(E,E_0)$ embodies the nonspatial 
information of the degradation process. It represents the equilibrium 
number of electrons per unit energy at an energy $E$ resulting from the
local energy degradation of an incident electron of energy $E_0$, and
can be used to calculate the yield $J_j$ of any state $j$ at energy 
$E_0$ with the help of following equation:

\begin{equation}
     J_j(E_0)=\int_{W_{th}}^{E_0} U(E,E_0)\: P_j(E)\, dE %eq 13
\end{equation}
where $P_j(E)=\sigma_j(E)/\sigma_T(E)$ is the probability of 
occurrence of the $j$th process whose threshold potential is $W_{th}$.
The yield for a particular process obtained by using the above equation
is used in the following sections to calculate the mean energy per ion 
pair and efficiencies for various loss processes. Except at very low 
energies, yield spectrum $U(E,E_0)$ and probability of excitation 
$P_j(E)$ both vary with $E$ in a much simpler manner than do $f(E,E_0)$
and $\sigma_j(E)$.  

For many application purposes yield spectrum obtained by equation (11) 
is represented in the following form:

\begin{equation}
      U(E,E_0)=U_a(E,E_0)\ H(E_0-E-E_m)+\delta(E_0-E).  %equation 14
\end{equation}
Here $H$ is the Heavyside function, with $E_m$ being the minimum 
threshold of the processes considered, and $\delta(E_0-E)$ is the 
Dirac delta function which allows for the contribution of the source 
itself.
In atmospheric and astrophysical applications it is convenient to
represent $U_a(E,E_0)$ in an analytical form [\textit{Green et al.}, 1977]:

\begin{equation}
      U_a(E,E_0)=A_1\xi _0^s+A_2(\xi _0^{1-t}/\epsilon^{3/2 +r}) %eq 15
\end{equation}
Here $\xi=E_0/1000$ and $\epsilon=E/I$ ($I$ is equal to lowest ionization 
threshold). $A_1=0.027,\ A_2=1.20,\ t=0,\ r=0,$ and $s=-0.0536$ are 
the best fit parameters.

We have also tried two other analytical forms given by  
\textit{Singhal et al.} [1980] and \textit{Green et al.} [1985]. 
The form given by \textit{Singhal et al.} [1980] is:

\begin{equation}
     U_a(E,E_0)=C_0+C_1\ \chi + C_2\ \chi^2    %eq 16
\end{equation}
Here $\chi=E_0^\Omega/(E+L)$; where $\Omega=0.585$ and $L=1.0$ and $E_0$ 
is in keV, $C_0=0.0185,\ C_1=5.98,$ and $C_2=210.4$ are fitted parameters.
The analytical form given by \textit{Green et al.} [1985] is:

\begin{equation}
      U_a(E,E_0)=C_0+C_1(E_k+K)/[(E-M)^2+L^2].  %equation 17
\end{equation}
Here $E_k=E_0/1000$, and $C_0$, $C_1$, $K$, $M$, and $L$ are the
fitted parameters which are independent of the energy. 
The values of these constant parameters 
are $C_0=0.0299 $, $C_1=430$, $K=0.0041$ keV, $M=0.31$ eV, and $L=1.9$ eV.

In obtaining our analytical fits we did not include values of the yield 
spectra very close to $E_0$ because in this regime yield spectra contain 
the rapid oscillations known as ``Lewis effect'' [cf. \textit{Douthat}, 
1975]. These oscillations are channels with a finite number of threshold 
energies, so that there are only certain energies near $E_0$ which an 
electron can acquire. Obviously, no electron can acquire an energy between 
$E_0$ and $E_0-E_m$, and that is why the Heavyside function $H$ is 
inserted in the first term on the right-hand side of equation (14).
The numerical yield spectrum represented analytically using
equations (15), (16), and (17) is the two-dimensional analytical yield 
spectrum (AYS). In our studies, we have used the AYS obtained using 
equation (15), which is presented in Figure 6 along with the numerical 
yield spectra obtained by using (14). 
It is clear from Figure 6 that the analytical spectra represents 
quite well the numerical yield spectra above the ionization threshold; 
however, at lower energies (below 15 eV) the AYS departs from the 
numerical yield spectra. Similar behavior is seen in the AYS of  
\textit{Green et al.} [1977]. 

To overcome this deficiency we introduce an additional function to
modify the lower energy part of the AYS:

\begin{equation}
        U_b(E,E_0)=\frac{E_0A_0e^{x}/A_1}{(1+e^{x})^2},      %eq 18
\end{equation}
Here $x=(E-A_2)/A_1$, and $A_0$, $A_1$, and $A_2$ are the fitting 
parameters. The values of parameters are $A_0=10.095$, $A_1=5.5$, 
and $A_2=0.9$.
Equation (18) only affects the lower energy ($\leq$15 eV) part of the fit. 
The final AYS is the sum of equations (15) and (18) which is shown in 
Figure 6 at several incident energies: depicting a better fit at lower 
energies ($>$5 eV) as well as at higher energies.

Because of the simplicity of function and cost effective computational 
advantage, the AYS technique has been widely used in different 
planetary atmospheres for various aeronomical calculations, like 
steady state electron fluxes and volume production rates for any 
ionization or excitation state; the details of the computational 
technique are described in earlier papers [e.g., \emph{Singhal 
and Haider}, 1984; \emph{Bhardwaj and Singhal}, 1993;
\emph{Singhal and Bhardwaj}, 1991; \emph{Bhardwaj et al.}, 1990, 1996;
\emph{Bhardwaj}, 1999, 2003; \emph{Bhardwaj and Michael}, 1999a, b; 
\emph{Michael and Bhardwaj}, 2000; \textit{Haider and Bhardwaj}, 2005].
               
\section{Mean Energy per Ion Pair}
The mean energy per ion pair, $\mu_j$, is defined as the incident
energy $E_0$ divided by the number of ion pairs produced. It can be
expressed as

\begin{equation}
     \mu_j(E_0)=E_0/J_j(E_0),            %eq 17
\end{equation}
where $J_j(E_0)$ is the population of the $j$th ionization process
obtained by equation (13). The quantity mean energy per ion pair is known 
to approach a constant value at higher energies.

Figure 7 shows the mean energy per ion pair for the ions 
CO$_2^+$ (including the ground and excited states), CO$^+$, O$^+$, C$^+$, 
CO$_2^{++}$, O$^{++}$ and C$^{++}$
along with the mean energy per ion pair for neutral \car, solid symbol  
represents the mean energy per ion pair for neutral \car\ obtained 
directly from the Monte Carlo simulation at few energy points. 

Mean energy for all the ions decreases very rapidly above their 
threshold value, but after $\sim$100 eV $\mu$ declines slowly   
and at higher energies it becomes almost constant.
The values of $\mu$ for CO$_2^+$, CO$^+$, O$^+$, and 
C$^+$ at 200 (1000) eV are 53.6 (51.2), 403 (415), 263.1 (247.8),
and 626.7 eV (576.2) eV, respectively. 
The mean energy per ion pair for neutral \car\ gas
is 37.5 (35.8) eV  at 200 (1000) eV. 
\emph{Fox and Dalgarno} [1979] reported a value of 33.1 eV at 
200 eV for the $\mu$, while \emph{Green et al.} [1977] obtained a value 
of 34.7 eV at 200 eV from their MDEB method. 
The measured value of the mean energy per ion pair in neutral \car\ 
is 32.7 at high energies [\textit{Klots}, 1968].
Mean energy per ion pair for X$^2\Pi_g$, A$^2\Pi_u$, B$^2\Sigma_u^+$,
and C$^2\Sigma_g^+$ states of CO$_2^+$ at 200 (1000) eV are 112.3 
(118.4), 180.3 (156), 301.5 (266.4), and 1999 (1222) eV,
respectively.

\section{Secondary Electron Distribution}
During the degradation process, every time the electron undergoes an 
ionization collision event, a secondary electron is produced. The energy 
of the secondary electron produced is calculated using (10). The maximum 
energy of the secondary electron produced can be $(E-I)/2$, where $E$ is 
the energy of the colliding electron and $I$ is the ionization potential. 
As mentioned before, secondary and tertiary electrons are also treated in 
the same manner as the primary electrons in the Monte Carlo model. The 
energy distribution of secondary electrons is presented in Figure 8 at 
several incident energies showing the number of secondary electrons 
produced per incident primary electron. The energy distributions of 
tertiary and quaternary electrons, which are presented only at E$_0=1000$ 
eV, are much steeper than that of secondary electrons.
Each incident electron of E$_0 = 1000$ eV, at some point of its 
energy degradation process, produces at least one secondary 
or tertiary or quaternary electron, whose energy is $<$7 eV.

\section{Efficiency}
As the electron collide with the atmospheric particles, they lose their
energy and finally become thermalized. The energy of the 
colliding electron is divided among the various inelastic loss 
processes. Efficiency  means the fraction of incident energy 
of the electron which is eventually deposited in a particular loss 
channel after the completion of the entire degradation process. The 
efficiency, $\eta_j(E_0)$, of the $j$th process at incident energy $E_0$ 
can be obtained as

\begin{equation}
      \eta_j(E_0)=\frac{W_{th}}{E_0}\; J_j(E_0)      %eq 18
\end{equation}
We have calculated the efficiencies for all inelastic collisions using 
numerical yield spectra obtained from equation (14) and the AYS
[sum of equations (15) and (18)].

Figure 9 presents efficiencies of various single ionization 
events producing CO$_2^+$, CO$^+$, O$^+$, and C$^+$. The CO$_2^+$ has 
the maximum efficiency throughout the energy region due its higher 
ionization cross section. At 1000 eV, $\sim$31\% energy of the incident 
electron goes into CO$_2^+$ formation, while 5.9\%, 9.8\%, 
and 5.0\% energy goes into the production of CO$^+$, O$^+$, and 
C$^+$, respectively. At higher energies ($>$100 eV), increase in the 
efficiencies for all ions is small, but near threshold it falls very 
rapidly. At threshold, efficiencies for CO$_2^+$, 
CO$^+$, O$^+$, and C$^+$ are 5.1\%, 1.1\%, 0.16\% and 0.19\%, 
respectively, while at 200 eV these are 29\%, 6.0\%, 9.2\%, and 4.6\%, 
respectively. Efficiencies for CO$_2^+$(A-X), CO$_2^+$(B-X), and first 
negative band of CO$^+$(B-X) are also shown in Figure 9. At 200 (1000) eV, 
12.2 (11.6)\% of incident electron energy goes in to the emission 
CO$_2^+$(A-X), while 9.8 (11.4)\% and 3.0 (3.3)\% goes in to the 
emissions CO$_2^+$(B-X) and CO$^+$(B-X), respectively.  

Figure 10 shows the efficiencies for double ionization of \car. At 
200 (1000) eV, efficiencies for CO$_2^{++}$, O$^{++}$, and C$^{++}$ 
are 0.56 (0.67)\%, 0.052 (0.12)\%, and 0.092 (0.14)\%, respectively.
We have also calculated the efficiencies for (CO$^+$,O$^+$), 
(C$^+$,O$^+$), and (O$^+$,O$^+$), based on cross sections of 
\emph{Tian and Vidal} [1998], whose values are 2.7 (3.1)\%, 1.8 
(2.4)\%, and 0.96 (1.1)\%  at 200 (1000) eV.
It is clear from Figures 9 and 10, that efficiencies calculated 
from the  model and those obtained by using AYS are in good agreement.

Efficiencies for various  excitation processes are presented in Figure 11. 
The 13.6, 12.4, and 11.1 eV states dominate the excitation events having 
efficiencies 16 (15)\%, 12 (13)\%, and 4.7 (4.2)\%
at 200 (1000) eV, respectively.
Efficiencies of various line emissions of atomic oxygen and carbon are 
shown in Figure 12. Efficiencies for O I (1304), O I (1356), C I (1279), 
C I (1329), C I (1561), and C I (1657), are 0.12 (0.13)\%, 0.27 (0.28)\%, 
0.084 (0.089)\%, 0.035 (0.030)\%, 0.10 (0.093)\%, and 0.19 (0.18)\%, 
respectively, at 200 (1000) eV. Overall efficiencies calculated from 
numerical yield spectra and AYS for various emission and excitation events 
are in good agreement.

In Figure 13, we present a summary picture of the electron energy 
distribution in \car\ for all the loss processes grouped into important 
loss channels. At higher ($>$50 eV) energies ionization is the dominant 
loss process with energy consumption of $\sim$50\%. 
At lower energies ($<$15 eV), 11.1, 12.4, 8.6, and 9.3 eV loss 
channels are more important. At energies below 10 eV, vibration 
becomes the main loss channel. We have also shown the efficiency  
for total attachment process, which produces negative ion O$^-$. 
The efficiency for anion O$^-$ production peaks around 8 eV  
with a value of 0.8\%, while it is 0.15 (0.13)\% at 200 (1000) eV.
The total efficiency for double ionization, which results in the 
production of CO$_2^{++}$, O$^{++}$, and C$^{++}$ ions, is also 
depicted in the figure. The double ionization efficiency raises sharply 
above 40 eV, having value of  0.4 (0.7)\%  at 100 (200) eV. Around 1000
eV, double ionization efficiency is  0.9\%, which is higher than 
that of 8.6 and 9.3 eV excitation states. On the other hand, at energies
$>$100 eV efficiency for dissociative ionization is higher than that of 
the 13.6 and 12.4 eV states.

\section{Summary}
In this paper we have presented a Monte Carlo model for 
$\le$1000 eV electron degradation in \car\ gas. All the  
e-\car\ collision cross sections are compiled and fitted analytically. 
The analytical cross sections are presented in  
figures along with the laboratory measured cross sections for direct 
comparison, and the fitting parameters are provided in tables.
The output of the Monte Carlo model is used to calculate the 
numerical ``yield spectra'', which is represented by an analytical form. 
This analytical yield spectra (AYS) can be used in planetary 
atmospheres to determine various aeronomical quantities. 
We have modified and improved the AYS presented by \textit{Green et al.} 
[1977] and \textit{Singhal et al.} [1980] by adding a term that provides 
a better analytical representation of yield spectra at lower ($<$15 eV) 
energies.
The yield spectra is employed to compute the mean energy per ion pair 
and efficiency of various inelastic processes. The mean energy per ion 
pair for  \car\ is found to be 37.5 (35.8)  at 200 (1000) eV.
The energy distribution of secondary electrons produced per incident 
electron is presented at few incident energies. 

Efficiency is an effective measure to know what
fraction  of the incoming particle energy  goes into a 
particular loss channel. We have presented efficiencies for various 
inelastic events calculated by using the AYS as well as by using the 
numerical yield spectra obtained from the Monte Carlo model. Efficiencies 
obtained by the two methods are in good agreement. In addition to major 
inelastic processes, efficiencies are presented for the formation of 
negative ions, double and dissociative double ionization of \car, and 
total vibrational excitation in the (100), (010), and (001) states.
Since the AYS do not represent well the numerical yield spectra at 
very low ($<$5 eV) energies, the yield for vibrational excitation
and attachment processes calculated by the AYS would be approximate.
Ionization is the dominant loss process at higher energies, above 100 eV 
$\sim$50\% energy goes into ionization.  
At energies around and below ionization threshold  excitation processes 
become important, and at energy below 10 eV, vibration is the dominant 
loss channel consuming more than 70\% energy. 
The  13.6 and 12.4 eV loss channels are also important, at 
1000 eV, around 28\% of incident particle energy goes in to these states. 
A part of these states represents the emissions of Cameron band system, 
which is an important emission in atmospheres of Mars and Venus as well 
as on comets (\textit{Bhardwaj and Raghuram}, 2009, in preparation). 

Efficiencies presented in this paper can be applied to planetary 
atmospheres by folding them with electron production rate and integrating 
over the energy. These results will be useful in the  modeling of  
aeronomical processes in atmospheres of Mars, Venus, and \car-containing
atmospheres.

\end{article}

\newpage
%% Table 1

\begin{table}
%\small
\caption{Elastic differential cross section for electron impact on 
\car\ (in units of $10^{-16}$ cm$^2$/sr)}
\begin{tabular}{lcccccccccc}
\hline\hline
\multicolumn{1}{c}{\centering Energy } & \multicolumn{10}{c}{Angle (degree)} \\ \cline{2-11}
  (eV)   & 0 & 5 & 10 &  15 & 20 & 30 & 40 & 50 & 60 & 70  \\[5pt]
\hline
1.5    & (1.350)\footnote & (1.252) & (1.154) & (1.056) & 0.9580 & 0.7620 & 0.5410 & 
       0.4050 & 0.3289 & 0.2957   \\[5pt]
2.0    & (1.157) & (1.055) & (0.954) & (0.852) & 0.7505 & 0.5472 & 0.3896 & 
       0.2455 & 0.2368 & 0.2489   \\[5pt]
3.0    & (1.174) & (1.060) & (0.945) & (0.831) & 0.7160 & 0.4868 & 0.3069 & 
       0.3118 & 0.3386 & 0.3779   \\[5pt]
3.8    & (2.295) & (2.059) & (1.824) & (1.589) & 1.3536 & 0.8831 & 0.6294 & 
       0.5897 & 0.5715 & 0.5367   \\[5pt]
4.0    & (2.007) & (1.844) & (1.681) & (1.517) & 1.3536 & 1.0269 & 0.7770 & 
       0.6857 & 0.6472 & 0.5834   \\[5pt]
5.0    & (0.250) & (0.333) & (0.416) & (0.499) & 0.5824 & 0.7486 & 0.8076 & 
       0.8994 & 0.8079 & 0.7272   \\[5pt]
6.0    & (0.501) & (0.546) & (0.592) & (0.637) & 0.6823 & 0.7730 & 0.8244 & 
       0.8383 & 0.8373 & 0.7644   \\[5pt]
6.5    & (0.933) & (0.914) & (0.896) & (0.878) & 0.8599 & 0.8236 & 0.9132 & 
       0.9286 & 0.8950 & 0.6978   \\[5pt]
7.0    & (0.986) & (0.960) & (0.934) & (0.909) & 0.8828 & 0.8313 & 0.9039 & 
       0.9391 & 0.8025 & 0.7558   \\[5pt]
8.0    & (16.88) & (12.95) & (9.02) & 5.0890 & 1.1590 & 1.0020 & 0.9640 & 
       0.8542 & 0.7214 & 0.6755   \\[5pt]
9.0    & (24.24) & (18.56) & (12.87) & 7.1830 & 1.4960 & 1.5880 & 1.0870 & 
       0.9487 & 0.8375 & 0.6622   \\[5pt]
10.0   & (39.19) & (29.96) & (20.74) & 11.520 & 2.2977 & 1.5342 & 1.2136 &
       0.9926 & 0.7430 & 0.6260   \\[5pt]
15.0   & (31.84) & (24.84) & (17.84) & 10.843 & 3.8430 & 2.7180 & 1.7789 & 
       1.1756 & 0.7997 & 0.5777   \\[5pt]
20.0   & (13.80) & (11.77) & (9.743) & 7.7149 & 5.6871 & 3.2623 & 1.8542 &
       1.2248 & 0.7475 & 0.4324   \\[5pt]
30.0   & (19.89) & (17.10) & (14.31) & (11.52) & 8.7310 & 3.1540& 1.4363&
       0.7430 & 0.4678 & 0.3060   \\[5pt]
40.0   & (15.70) & (13.51) & (11.31) & (9.115) & 6.9200 & 2.5300 & 1.0400 &
       0.5300 & 0.3100 & 0.1800   \\[5pt]
50.0   & 14.820 & 12.690 & 10.560 & 8.4300 & 6.3000 & 2.0400 & 0.8100 &
       0.4000 & 0.2100 & 0.1440   \\[5pt]
60.0   & (13.44) & (11.50) & (9.556) & (7.614) & 5.6710 & 1.7860 & 0.6597 &
       0.3412 & 0.1683 & 0.1109   \\[5pt]
70.0   & (10.61) & (9.055) & (7.50) & (5.945) & 4.3900 & 1.2800 & 0.5200 &
       0.2500 & 0.1420 & 0.1130   \\[5pt]
80.0   & (9.79) & (8.35) & (6.91) & (5.470) & 4.0300 & 1.1500 & 0.4700 &
       0.2200 & 0.1360 & 0.1090   \\[5pt]
90.0   & (8.50) & (7.24) & (5.98) & (4.72) & 3.4600 & 0.9400 & 0.3800 &
       0.2000 & 0.1450 & 0.1120   \\[5pt]
100.0  & (9.273) & (7.893) & (6.514) & (5.134) & 3.7543 & 0.9950 & 0.3969 &
       0.2026 & 0.1502 & 0.1124   \\[5pt]
200.0  & (31.75) & (22.68) & 13.610 & 4.5390 & 2.4170 & 0.6160 & 0.3380 &
       0.2230 & 0.1270 & 0.0952   \\[5pt]
300.0  & (19.35) & (13.86) & 8.3720 & 2.8850 & 1.2350 & 0.3880 & 0.2670 &
       0.1290 & 0.0716 & 0.0539   \\[5pt]
400.0  & (16.82) & (12.00) & 7.1900 & 2.3770 & 1.0400 & 0.4550 & 0.2150 &
       0.0968 & 0.0624 & 0.0523   \\[5pt]
500.0  & (132.80) & 77.22 & 21.600 & 6.6100 & 2.7800 & 1.5600 & 0.7130 &
       0.3140 & 0.2190 & 0.1620   \\[5pt]
800.0  & (138.20) & 75.020 & 11.810 & 4.1800 & 2.6500 & 0.9050 & 0.3280 &
       0.1900 & 0.0920 & 0.0673   \\[5pt]
1000.0 & (113.0) & 62.100 & 11.200 & 3.5500 & 2.3500 & 0.6600 & 0.2820 &
       0.1430 & 0.0925 & 0.0640   \\[5pt]
\hline
\end{tabular}
\end{table}
%\end{center}
%% Enter Figures and Tables at here:
\addtocounter{table}{-1}
%\begin{center}
\begin{table}
%\small
\caption{Contd.}
\begin{tabular}{lcccccccccc}
\hline\hline
\multicolumn{1}{c}{\centering Energy } & \multicolumn{10}{c}{Angle (degree)} \\ \cline{2-11}
(eV)& 80 & 90 & 100 &  110 & 120 & 130 & 135 & 150 & 165 & 180  \\[5pt]
\hline
1.5    & 0.2700 & 0.2405 & 0.3080 & 0.3040 & 0.3567 & 0.3650 & (0.3629)
 &     (0.3816) & (0.3941) & (0.4065)   \\[5pt]
2.0    & 0.2765 & 0.2845 & 0.3021 & 0.3276 & 0.3776 & 0.3992 & (0.4100)
 &     (0.4424) & (0.4748) & (0.5072)   \\[5pt]
3.0    & 0.3876 & 0.3937 & 0.3950 & 0.4380 & 0.4830 & 0.5173 & (0.5345)
 &     (0.5859) & (0.6374) & (0.6888)   \\[5pt]
3.8    & 0.5539 & 0.5739 & 0.5096 & 0.5187 & 0.5280 & 0.5475 & (0.5573)
 &     (0.5865) & (0.6158) & (0.6450)   \\[5pt]
4.0    & 0.5595 & 0.5037 & 0.4431 & 0.4217 & 0.4258 & 0.4803 & (0.5076)
 &     (0.5893) & (0.6711) & (0.7528)   \\[5pt]
5.0    & 0.6026 & 0.4794 & 0.3910 & 0.2647 & 0.2523 & 0.2853 & (0.3018)
 &     (0.3513) & (0.4008) & (0.4503)   \\[5pt]
6.0    & 0.6422 & 0.5258 & 0.4518 & 0.3476 & 0.3136 & 0.3798 & (0.4129)
 &     (0.5122) & (0.6115) & (0.7108)   \\[5pt]
6.5    & 0.6616 & 0.5300 & 0.4252 & 0.3390 & 0.3201 & 0.3520 & (0.3680)
 &     (0.4158) & (0.4637) & (0.5115)   \\[5pt]
7.0    & 0.6258 & 0.5273 & 0.4333 & 0.3766 & 0.3798 & 0.3724 & (0.3687)
 &     (0.3576) & (0.3465) & (0.3354)   \\[5pt]
8.0    & 0.6761 & 0.5343 & 0.4596 & 0.4263 & 0.4058 & 0.5183 & (0.5746)
 &     (0.7433) & (0.9121) & (1.0810)   \\[5pt]
9.0    & 0.5799 & 0.5394 & 0.4811 & 0.4381 & 0.4816 & 0.6006 & (0.6601)
 &     (0.8386) & (1.0170) & (1.1960)   \\[5pt]
10.0   & 0.5468 & 0.4856 & 0.4478 & 0.4319 & 0.5304 & 0.7077 & (0.7964)
 &     (1.0620) & (1.3280) & (1.5940)   \\[5pt]
15.0   & 0.4471 & 0.3596 & 0.3673 & 0.4046 & 0.5445 & 0.7832 & (0.9026)
 &     (1.2610) & (1.6190) & (1.9770)   \\[5pt]
20.0   & 0.3516 & 0.3041 & 0.3071 & 0.3887 & 0.5493 & 0.6738 & (0.7361)
 &     (0.9228) & (1.1100) & (1.2960)   \\[5pt]
30.0   & 0.1896 & 0.1882 & 0.2391 & 0.2536 & 0.3195 & 0.4441 & (0.5064)
 &     (0.6933) & (0.8802) & (1.0670)   \\[5pt]
40.0   & 0.1330 & 0.1190 & 0.1130 & 0.1500 & 0.2400 & (0.330) & (0.3750)
 &     (0.5100) & (0.6450) & (0.7800)   \\[5pt]
50.0   & 0.1180 & 0.0920 & 0.0810 & 0.1300 & 0.2500 & (0.370) & (0.4300)
 &     (0.6100) & (0.7900) & (0.9700)   \\[5pt]
60.0   & 0.0936 & 0.0911 & 0.0812 & 0.01175& 0.1805 & 0.2748 & (0.3220)
 &     (0.4634) & (0.6049) & (0.7463)   \\[5pt]
70.0   & 0.1040 & 0.0850 & 0.0910 & 0.1400 & 0.2100 & (0.280) & (0.3150)
 &     (0.4200) & (0.5250) & (0.6300)   \\[5pt]
80.0   & 0.0900 & 0.0850 & 0.0900 & 0.1200 & 0.1800 & (0.240) & (0.2700)
 &     (0.3600) & (0.4500) & (0.5400)   \\[5pt]
90.0   & 0.0804 & 0.0890 & 0.0920 & 0.1200 & 0.1600 & (0.200) & (0.2200)
 &     (0.2800) & (0.3400) & (0.4000)   \\[5pt]
100.0  & 0.0840 & 0.0697 & 0.0754 & 0.0880 & 0.1076 & 0.1373 & (0.1522)
 &     (0.1967) & (0.2413) & (0.2858)   \\[5pt]
200.0  & 0.0756 & 0.0646 & 0.0709 & 0.0770 & 0.0804 & 0.0878 & (0.0944) &
       (0.1142) & (0.1340) & (0.1538)   \\[5pt]
300.0  & 0.0505 & 0.0376 & 0.0337 & 0.0314 & 0.0272 & 0.0233 & (0.0217) &
       (0.0169) & (0.0121) & (0.0073)   \\[5pt]
400.0  & 0.0376 & 0.0305 & 0.0256 & 0.0255 & 0.0236 & 0.0223 & (0.0202) &
       (0.0139) & (0.0076) & (0.0013)  \\[5pt]
500.0  & 0.1080 & 0.0843 & 0.0752 &0.0658&0.0548&(0.0438)&(0.0383)&
       (0.0218) & (0.0053) & (0.000012) \\[5pt]
800.0  & 0.0523 & 0.0319 & 0.0283 & 0.0238 & 0.0221 & (0.0204) &(0.0195)&
       (0.0170) & (0.0145)& (0.01190)  \\[5pt]
1000.0 & 0.0360 & 0.0275 & 0.0220 & 0.0165 & 0.0149 & (0.0133)&(0.0125)&
       (0.0101) & (0.0077)& (0.0053)  \\[5pt]
\hline \\ 
\end{tabular}
\\
\footnotemark[1]{\small Values inside the bracket
indicates a linearly extrapolated value.}
\end{table}

%Table 2
\begin{table}
\caption{Parameters for Electron attachment process}
\begin{tabular}{lcccc}
\hline\hline
\multicolumn{4}{c}{}\\
 \quad & \quad $W_P$ \quad  & \quad $A$\quad  & \quad $U$ \quad   & \quad $W_{th}$\quad \\ 
\hline                                                    
$O^{-}$I   & \quad 4.3 \quad  &\quad $0.0013\times10^{-16}$\quad  & 
\quad 0.22 \quad   &  \quad 3.4 \quad      \\ [5pt]
$O^{-}$II  & \quad 8.1 \quad  &\quad $0.0056\times10^{-16}$\quad  &  
\quad 0.33 \quad   &  \quad 5.9 \quad   \\
\hline
\end{tabular}
\end{table}

%Table 3
\begin{table}
\caption{Parameters for various ionization processes}
\begin{tabular}{cccccccccccc}
\hline\hline
%multicolumn{1}{c}{} & \multicolumn{4}{c}{Single Ionization} & 
%multicolumn{7}{c}{\centering Double Ionization} \\ 
%cline{2-5} \cline{7-12} \\
      & W$_{th}$ & I & K & K$_B$  & J & J$_B$ & 
$\Gamma_S$ & $\Gamma_B$ & T$_S$ & T$_A$ & T$_B$ \\
\hline
CO$_2^+$(Total)&13.76 & 13.76 & 9.83 & 0.0&40.59 & 1.050 & 18.61 &-13.23 & 
-0.847 & 875  & 44.52  \\ [5pt]
CO$_2^+$(X$^2\Pi_g$) & 13.76 & 13.76 & 3.480 & 0.0 &4.099 &-2.35 & 11.11 & 
-13.26 & -0.847 & 1000 & 27.52 \\ [5pt]
CO$_2^+$(A$^2\Pi_u$) & 17.8 & 17.8 & 8.632 & 6.0 &86.36 & 1.004& 12.00 & 
-18.80 & -1.996 & 550& 10.20 \\ [5pt]
CO$_2^+$(B$^2\Sigma_u^+$)&18.1 &18.1 & 4.632& 5.0 & 85.36 & 1.004 &12.00  
& -18.5 & -0.978 & 450   &   10.20 \\ [5pt]
CO$_2^+$(C$^2\Sigma_g^+$)&19.4 &19.4 &0.580&0.0&21.19&1.270&10.98&-19.00& 
-0.887 & 1000 & 38.80    \\ [5pt]
 CO$^+$ & 24.76 & 24.76 & 1.347 & 15.00 &6.650 & 1.256 & 9.556 & -24.0 & 
1.887 &  800 & 25.52     \\ [5pt]
O$^+$ & 24.5 & 24.5 & 2.399 & 40.00 &11.34 &-1.10 &13.42 & -24.0 & 
-0.587 & 0.0 & 0.0 \\ [5pt]
C$^+$ & 29.5 & 29.5 & 1.659 & 0.0 &53.54 & 0.625 & 10.62 & -29.0 &
 2.800 & 21.8 & 44.00 \\ [5pt]
C$^{++}$ & 79.94 & 79.94 & 0.0012 & 25.00 &0.100 & 0.656 & 79.11 & 0.0
& 0.8473  & 0.0 & 0.0  \\ [5pt]
O$^{++}$ & 94 & 94 & 0.0055 & 10.00 &1.700 &-3.156 &20.11&0.0 &
-0.8473 & 0.0 & 0.0 \\ [5pt]
CO$_2^{++}$ & 37.6 & 44.75 & 0.0583 & 1.20 & 0.250 & 0.0 & 11.57 & 35.26 & 
1.548 & 800 & 30.52  \\[5pt] 
(CO$^+$,O$^+$)& 44.7 & 44.7 & 0.285 &1.200& 0.550 & 0.0 & 11.97 & 35.26 & 
1.547 & 650 & 30.52  \\ [5pt]
(C$^+$,O$^+$) & 44.7 & 44.7 & 0.288 & 1.200 & 0.550 & 0.0 & 10.40 & 
45.00 & 1.547 &750 & 0.0  \\ [5pt]
(O$^+$,O$^+$) & 44.7 & 44.7 &0.158 &1.200 &0.450  & 0.0 & 5.400 & 15.00 &
 5.547 &750 & 0.0  \\ 
\hline
\end{tabular}
\end{table}

%Table 4
\begin{table}
\caption{Parameters for various excitation and emission processes} 
\begin{tabular}[width=\textwidth]{lccccccc}
%\begin{tabular}{lccccccc}
\hline\hline
\multicolumn{4}{c}{}   & \multicolumn{4}{c}{} \\
Excitation states  &   $W$    &    $\alpha$  &  $\beta$  &  $W_J$   &  $\Omega$   & $F$   &    $A.F.$   \\
\hline
[Vibration, (010)]\footnotemark[1] & \,\ 0.080 \,\ &  \,\ 2.750 \,\ & \,\  1.000 \,\  & \,\  0.080 \,\ & \,\  0.750 \,\  &  \,\ 0.000060 \,\  & \,\,  0.0 \,\, \\[5pt]
[Vibration, (100)] &   0.180       &  1.070             &   1.000         &    0.180       &   0.750           &  0.000031   &   0.0     \\[5pt]
[Vibration, (001)] &   0.290       &  2.910             &   0.500         &    0.300       &   0.810           &  0.000445   &   0.0     \\[5pt]
8.6 eV state      &   8.600        &  0.556             &   2.000         &    8.600       &   0.936           &  0.060600   &   0.0     \\[5pt]
9.3 eV state      &   9.300        &  0.603             &   2.000         &    9.300       &   0.909           &  0.064000   &   0.0     \\[5pt]
11.1 eV state     &   7.760        &  0.246             &   3.000         &    11.100      &   1.110           &  4.420000   &   0.0     \\[5pt]
[12.4 eV state]   &   9.610        &  0.338             &   3.000         &    12.400      &   0.830           &  6.700000   &   0.0     \\[5pt]
[13.6 eV state]   &   10.50        &  0.625             &   3.000         &    13.600      &   0.849           &  3.350000   &   0.0     \\[5pt]
15.5 eV state     &   15.50        &  0.739             &   2.000         &    15.500      &   0.793           &  0.139000   &   0.750   \\[5pt]
16.3 eV state     &   12.30        &  0.605             &   3.000         &    16.300      &   0.911           &  0.716000   &   0.750   \\[5pt]
17.0 eV state     &   13.00        &  0.649             &   3.000         &    17.000      &   0.878           &  0.114000   &   0.750   \\[5pt]
17.8 eV state     &   14.00        &  0.977             &   3.000         &    17.800      &   0.725           &  0.051100   &   0.750   \\[5pt]
OI (1304)         &   20.10        &  0.599             &   3.000         &    22.000      &   1.000           &  0.127000   &   0.750   \\[5pt]
OI (1356)         &   16.40        &  0.600             &   3.000         &    20.400      &   0.944           &  0.168000   &   0.500   \\[5pt]
CI (1279)         &   15.70        &  1.000             &   3.000         &    26.200      &   0.643           &  0.010400   &   0.500   \\[5pt]
CI (1329)         &   21.80        &  1.000             &   3.000         &    20.900      &   1.040           &  0.020200   &   0.500   \\[5pt]
CI (1561)         &   22.40        &  1.000             &   3.000         &    24.500      &   0.982           &  0.053800   &   0.500   \\[5pt]
CI (1657)         &   21.10        &  1.000             &   3.000         &    24.100      &   0.947           &  0.872000   &   0.500   \\[5pt]
[CO$^+$(first negative)]& 18.13 & 0.656 & 2.54 & 25.11 & 0.804 & 
1.055 & 0.0 \\  
\hline \\
\end{tabular} 
\\
\footnotemark[1]{\small{Parameters are taken from \textit{Jackman et al.} 
[1977], except for the states which are \\ inside the square brackets whose 
parameters have been  modified.}}
\end{table}

\clearpage
%\newpage

%Figure 1
\begin{figure}
\centering
\noindent\includegraphics[width=20pc]{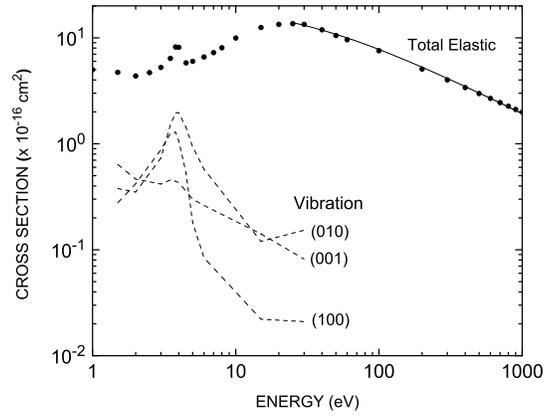}
\caption{The e-\car\ total elastic cross section and vibrational excitation
cross sections for three modes. For elastic cross section, symbol 
represents the cross section values of \emph{Itikawa} [2002], and solid 
curve represents the analytical fit using equation (1). Dashed curve 
represents the vibrational excitation cross sections taken from 
\emph{Itikawa} [2002]}
\end{figure}

%Figure 2
\begin{figure}
\centering
\noindent\includegraphics[width=20pc]{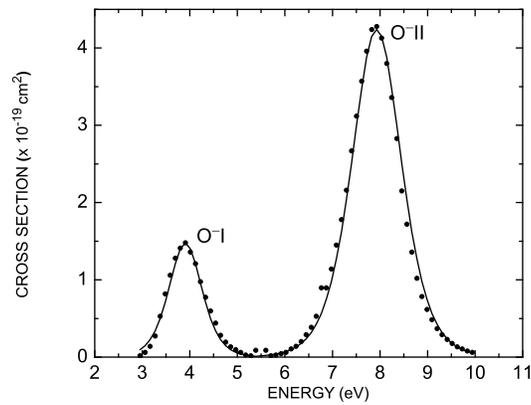}
\caption{Dissociative electron attachment cross section for the formation 
of O$^-$ ion. Symbol represents the values of \emph{Itikawa} 
[2002] based on \emph{Rapp and Briglia} [1965]; solid curve represents 
analytical fit of  O$^-$ cross section using equation(2). I and II 
denotes the first and second peak.}
\end{figure}

%Figure 3
\begin{figure}
\centering
\noindent\includegraphics[width=20pc]{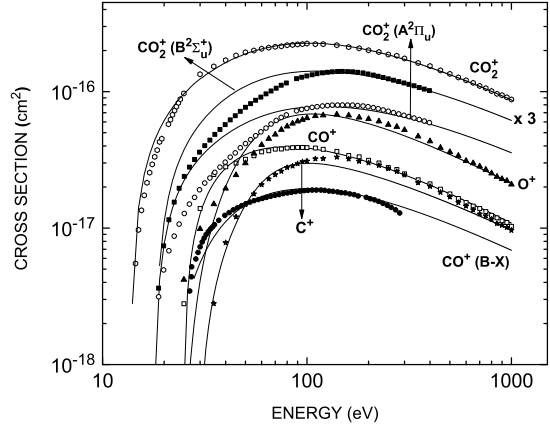}
\caption{Ionization and emission cross sections of \car. Symbol 
represents the  values of \emph{Itikawa} [2002], and solid curve 
represents the analytical fits using equation (3) except for CO$^+$(B-X) 
state, which is fitted using equation (4). Note that the cross section for 
CO$_2^+$(B$^2\Sigma_u^+$) state has been plotted after multiplying by a factor 
of 3.}
\end{figure}

%Figure 4
\begin{figure}
\centering
\noindent\includegraphics[width=20pc]{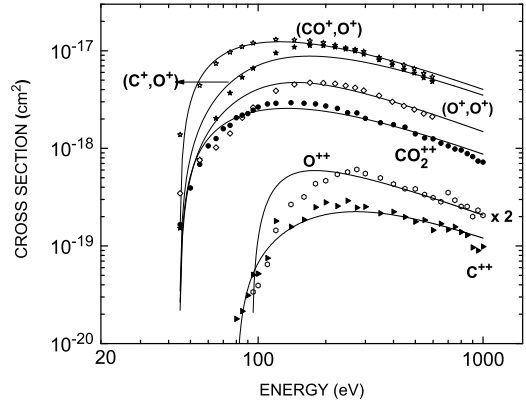}
\caption{Cross sections for electron impact double ionization of \car. 
Symbol represents the measured cross section values and solid 
curve represents the analytical fit using equation (3).
Cross sections for (CO$^+$,O$^+$), (C$^+$,O$^+$), and (O$^+$,O$^+$) 
have been taken from \emph{Tian and Vidal} [1998], and that for  
CO$^{++}_2$, O$^{++}$, and C$^{++}$ from \textit{Itikawa} [2002]. 
Cross section for O$^{++}$ has been plotted after multiplying by a 
factor of 2.}
\end{figure}  

%Figure 5
\begin{figure}
\centering
\noindent\includegraphics[width=20pc]{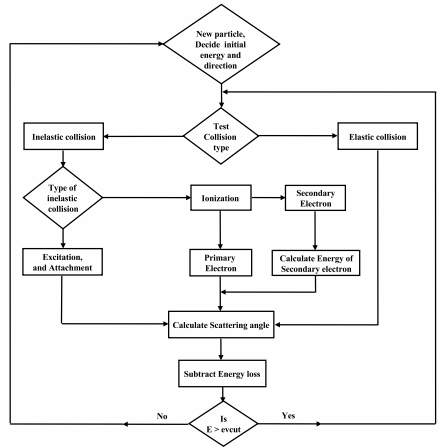}
\caption{A simplified flow diagram of the  Monte Carlo simulation. The 
diagram shows flow upto secondary electron, but tertiary and subsequent
electrons are also treated in a similar manner in the simulation.}
\end{figure}

%Figure 6
\begin{figure}
\centering
\noindent\includegraphics[width=20pc]{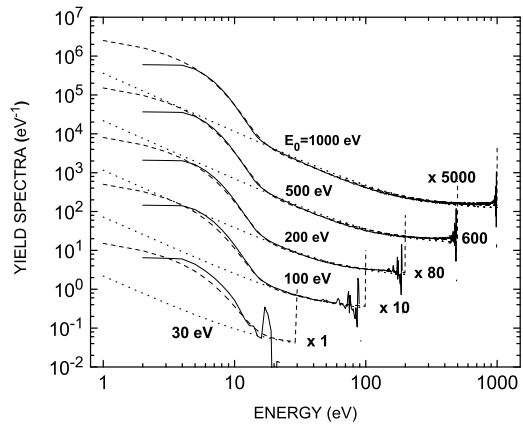}
\caption{The numerical yield spectra from the Monte Carlo model (solid 
curve) and AYS using equation (15)(dotted curve) at incident energies 
($E_0$) 30, 100, 200, 500,  and 1000 eV. Dashed curve represents the 
improved AYS calculated by summing equations (15) and (18). 
The yield spectra at 100, 200, 500, and 1000 eV are plotted after 
multiplying by a factor of 10, 80, 600, and 5000, respectively.}
\end{figure}

%Figure 7
\begin{figure}
\centering
\noindent\includegraphics[width=20pc]{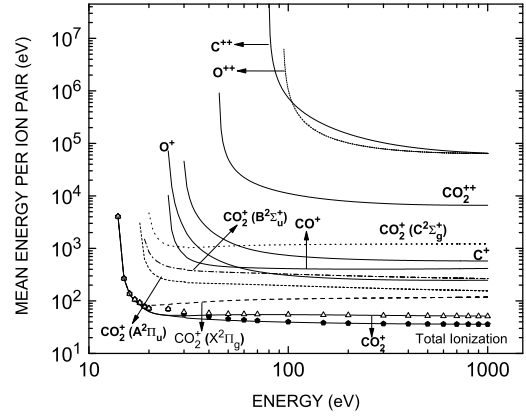}
\caption{The mean energy per ion pair for ions CO$_2^{+}$ (CO$_2^{+}$ is 
the sum of four states X$^2\Pi_g$,  A$^2\Pi_u$,  B$^2\Sigma_u^+$, and 
 C$^2\Sigma_g^+$), CO$^+$, O$^+$, C$^+$, CO$_2^{++}$, O$^{++}$, and 
C$^{++}$, and the neutral CO$_2$ gas (total), symbol represents the $\mu$ 
calculated using numerical yield spectra for the CO$_2^+$ and neutral 
\car.}
\end{figure}

%Figure 8
\begin{figure}
\centering
\noindent\includegraphics[width=20pc]{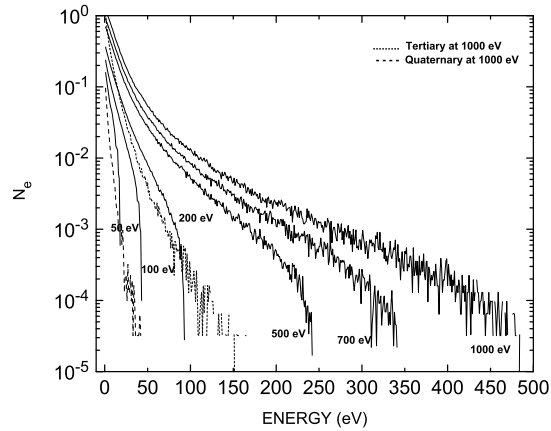}
\caption{The energy distribution of secondary electrons at six 
incident energies ($E_0$): 50, 100, 200, 500, 700, and 1000 eV.
$N_e$ represents the number of secondary, tertiary, or quaternary 
electrons produced per incident primary electron. Dotted curve and dashed 
curve represent energy distribution of tertiary and quaternary electrons, 
repectively at $E_0=1000$ eV.}
\end{figure}

%Figure 9
\begin{figure}
\centering
\noindent\includegraphics[width=20pc]{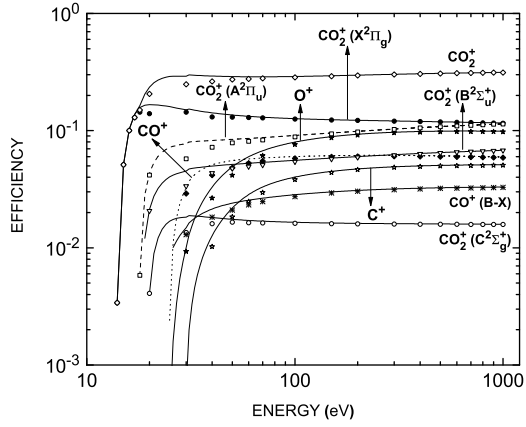}
\caption{Efficiencies of various ionization and emission processes.
Symbol represents the efficiency calculated by numerical yield spectra 
and curves represent the efficiency calculated by the AYS. 
CO$_2^+$(A$^2\Pi_u$) and CO$_2^+$(B$^2\Sigma_u^+$) represent FDB 
($A^2\Pi_u \rightarrow X^2\Pi_g$) and 
ultraviolet doblet ($B^2\Sigma_u^+\rightarrow X^2\Pi_g$) emissions, 
respectively, and CO$^+$(B-X) represents first negative band emission of 
CO$^+$ ion.}
\end{figure}

%Figure 10
\begin{figure}
\centering
\noindent\includegraphics[width=20pc]{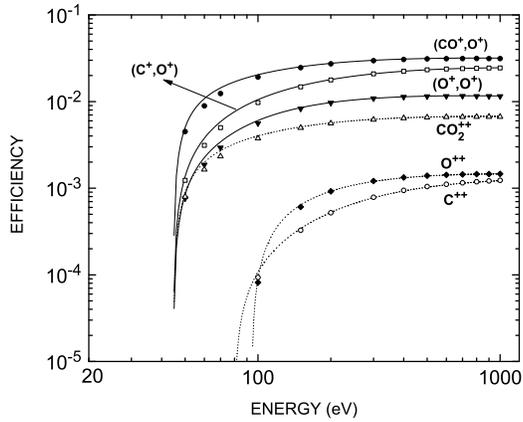}
\caption{Efficiencies for double ionization of \car\ due to electron 
impact. Symbol represents the efficiency calculated by numerical yield 
spectra and curves represent the efficiency calculated 
by the AYS.}
\end{figure}

%Figure 11 
\begin{figure}
\centering
\noindent\includegraphics[width=20pc]{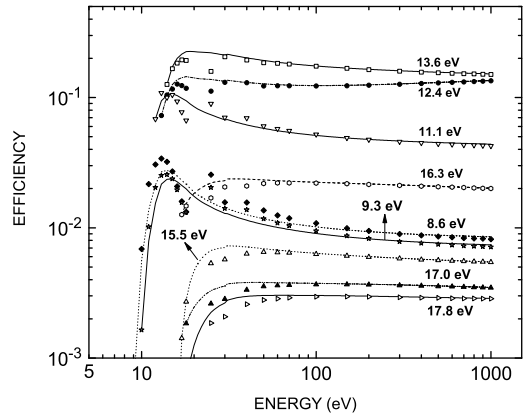}
\caption{Efficiencies of various excited states.
Symbol represents the efficiency calculated by numerical yield spectra 
and curves represent efficiency calculated by the AYS.}
\end{figure}

%\suppressfloats

%Figure 12 
\begin{figure}
\centering
\noindent\includegraphics[width=20pc]{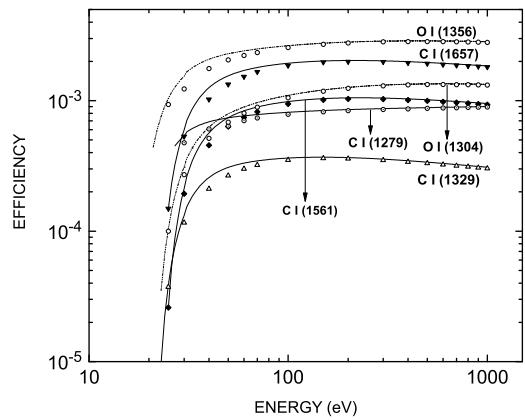}
\caption{Efficiencies of various oxygen and carbon line emissions.
Symbol represents the efficiency calculated by numerical yield spectra 
and curves represent the efficiency calculated by the AYS.}
\end{figure}

%Figure 13  
\begin{figure}
\centering
\noindent\includegraphics[width=20pc]{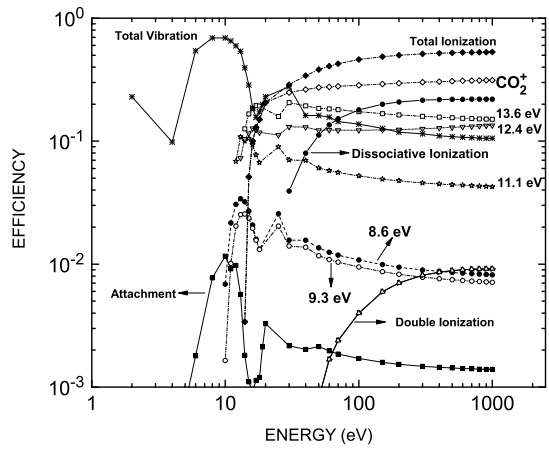}
\caption{Efficiencies for various important loss channels calculated using 
numerical yield spectra. Dissociative ionization includes the production 
of O$^+$, C$^+$, and CO$^+$ ions; double ionization includes the 
production of CO$_2^{++}$, O$^{++}$, and C$^{++}$ ions; and attachment 
denotes the production of O$^-$ ion.} 
\end{figure}

%\begin{center}

% When submitting articles through the GEMS system:
% COMMENT OUT ANY COMMANDS THAT INCLUDE GRAPHICS.

% Figure captions go below this illustration; Table captions above tables

% ONE-COLUMN figure/table, including eps graphics
%
% \begin{figure}
% \noindent\includegraphics[width=39pc]{samplefigure.eps}
% \caption{Caption text here}
% \end{figure}
% \end{document}
%
% \begin{table}
% \caption{}
% \end{table}
%
% ---------------
% TWO-COLUMN figure/table
%
% \begin{figure*}
% \noindent\includegraphics[width=39pc]{samplefigure.eps}
% \caption{Caption text here}
% \end{figure*}
%
% \begin{table*}
% \caption{Caption text here}
% \end{table*}
%
% see below for how to make landscape figures or tables

%%% End the article here:

\end{document}